\documentclass[12pt]{article}
\usepackage{epsfig}
\begin{document}
\topmargin .5cm
\begin{center}
{\bf {\Large Unusual modes and photonic gaps in a Vicsek waveguide network}}
\end{center}
\begin{center}
Sheelan Sengupta and Arunava Chakrabarti\\
Department of Physics, University of Kalyani, West Bengal 741 235, India.\\
\end{center}
\begin{abstract}
We propose a simple model of a waveguide network designed following the growth
rule of a Vicsek fractal. We show, within the framework of real space 
renormalization group (RSRG) method, that such a design may lead to the
 appearance of unusual electromagnetic modes. Such modes exhibit an extended
 character in RSRG sense. However, they lead to a power law decay in the
 end-to-end transmission of light across such a network model as the size of
 the network increases. This, to our mind, may lead to an observation of power
 law localization of light in a fractal waveguide network. The general
 occurence of photonic band gaps and their change as a function of the
 parameters of the system are also discussed.
\end{abstract}
\section{Introduction}
Propagation of electro-magnetic waves in dielectric materials has received
 extensive attention in recent years \cite{cms95}-\cite{dsw97}. The existence
 of photonic band gaps (PBG) in the transmission spectra of such materials is
 of particular interest and is of importance in both fundamental science and
 technological applications. These studies constitute principal part of
 mesoscopic physics, a major focus being the possibility of localization of
 light \cite{dzz94,slm91}, a direct evidence of which has already been reported
 for a strongly scattering media of semiconductor powders \cite{dsw97}. The
 localization of classical waves is purely a result of multiple scattering in a
 random environment and free from the complications arising from interaction
 effects. Though, due to Rayleigh scattering, it is more difficult to localize
 classical waves compared to electrons.
\vskip .2in
An alternative to the usual PBG systems may be obtained by forming network
with slender waveguide tubes. This geometry does not need materials with high
dielectric constants and yet, is capable of demonstrating Anderson localization
of light. Zhang et al \cite{zqz98} have experimentally observed Anderson
 localization of light in a three dimensional network consisting of nearly one
dimensional segments of waveguide. Vesseur et al \cite{jov99} have examined
 the photonic band structure of a comb-like waveguide geometry. This work
 provides an example of a one dimensional photonic crystal enabling one to
 investigate the occurrence of localized states in such systems. The network
 system can produce large gaps even in one dimension, though it may be 
sensitive to the structure of the unit cell. Recently, network models using
 serial loop structure have been studied in the context of propagation of
electro-magnetic and acoustic waves \cite{am03,aa04}. Such theoretical
 investigations are additional justified as the present day nano-technology
produce tailor-made waveguide networks with different geometries.
\vskip .2in
Photonic band gaps also exist in the network systems without periodicity, such
 as fractal networks \cite{ml00}. Due to their self-similar structures and the
absence of translational invariance within the structures, fractals 
\cite{ed83,rr82} are much more complicated to study compared to the periodic
 systems. Investigation of the propagation of classical waves in a network
 model designed following a fractal geometry offers an interesting opportunity
 to observe the interplay of deterministic but non-translational order and
 interference effects. Unfortunately, this aspect has been grossly overlooked
 in the literature except for the recent work done by Li et al \cite{ml00}, on a Sierpinski gasket network.
\vskip .2in
In the present work, we propose a model waveguide network based on the geometry
of a Vicsek fractal \cite{csj93}-\cite{jqy93}. We study the transmission of 
electromagnetic waves through finite sized Vicsek Waveguide Network (VWN) using
 real space renormalization group (RSRG) and transfer matrix (TM) methods. A
 regular Vicsek fractal is already known to possess exotic spectral properties
 \cite{csj93}-\cite{jqy93}. In addition to this, it has also been seen shown
 that an infinite Vicsek fractal can support a countable infinity of extended
 wavefunction in the context of electronic transport \cite{ac96,acb96}.
 Some of these states are `anomalous' in the sense that the behaviour of the
 RSRG flow of the parameters of the system and the electronic transport are
 incompatible to each other \cite{acb96}. This acts as an additional motivation
for our work. We wish to examine the appearence of photonic gaps in such a
 network system, the dependence of the gaps on the system size and the possible
 occurence of extended electromagnetic modes in such a deterministic fractal
 network.
\vskip .2in
We find interesting spectal behaviour and formation of photonic gaps in a VWN.
Additionally, we have been able to discern a number of eigenmodes in a VWN,
 some of which display a completely extended nature. For this latter set of
 states, the RSRG flow of the parameters gives a signature of extendedness,
 while the end-to-end light transmission decays following a power law as a
 function of the size of the network. This, to our mind,  may lead to a
 possible power law trapping of light from an experimental point of view. We
 present out results below.
\section{The model and the method}
\begin{figure}
\centering{\psfig{figure=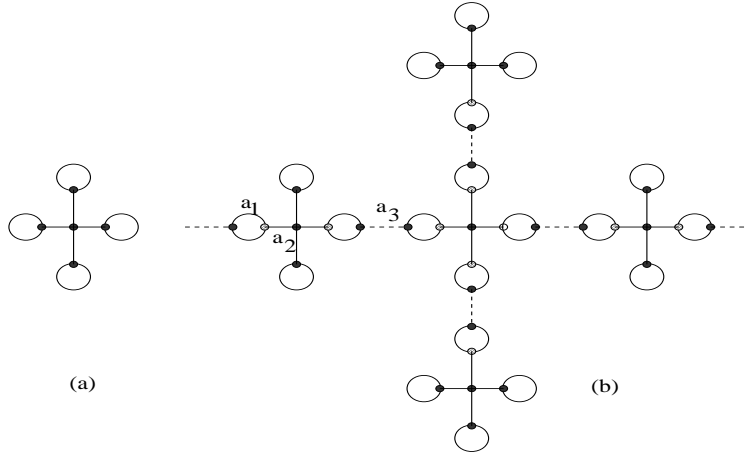,height=6cm,width=10cm}}
\caption{Vicsek waveguide network in the (a) first and (b) second generation.
The loops are attached to the open ends of a unit `cross' as shown in (a).
The patterned circles represent the nodes which are `decimated' to obtain a
Vicsek lattice of points.}\label{fig1}
\end{figure}

\begin{figure}
\centering{\psfig{figure=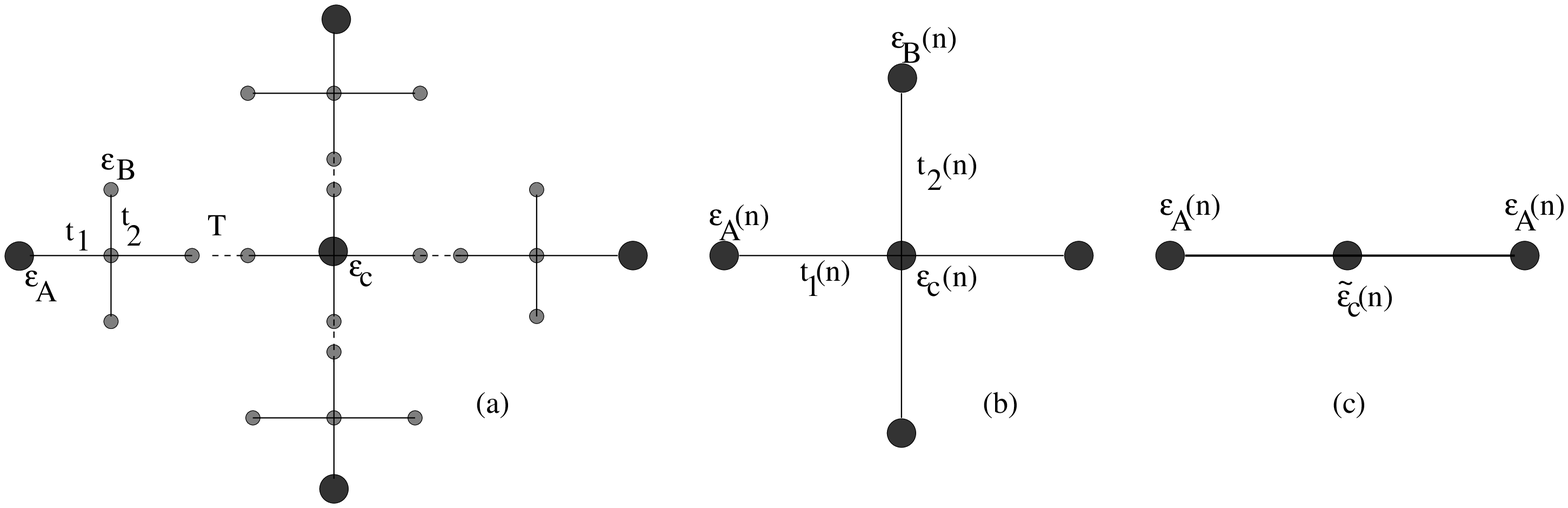,height=4cm,width=14cm}}
\caption{(a) A second generation VWN transformed into a Vicsek lattice of
 points on elimination of the patterned circles in Fig \ref{fig1}. The
 effective (renormalized) on-site potentials and hopping integrals are shown. 
(b) The one-step renormalized lattice obtained from (a). (c) The final
 triatomic molecule to be placed between the leads.} \label{fig2}
\end{figure}

A `unit cell' of a simple Vicsek geometry is obtained by considering a `cross'
 with four dangling ends \cite{jqy93} joined at a node, as a building unit.
The next generation fractal is obtained by placing four such `crosses' around a
 central cross and so on \cite{jqy93}. A VWN may be constructed by considering
 each arm of a `unit' cross to be consisting of a one dimensional waveguide
 segment. The dangling edges cause problems in realizing the boundary
 conditions from the standpoint of experiments. So, we attach loops of the same
 waveguide material at each end of a dangling arm. The first and the second
 generation VWN thus resemble what have been illustrated in Fig. \ref{fig1}.
 Each arm of a loop is of length $a_1$. The other segments are of lengths $a_2$
 and $a_3$ as shown in Fig. \ref{fig1}.
\vskip .2in
Following Zhang and Sheng \cite{zqz94} the `wave-function' $\psi_{i,i+1}$
 within any segment of length $l_{i,i+1}$ between the nodes $i$ and $i+1$ is
 given by,
\begin{eqnarray}
\psi_{i,i+1}(x) & = & \psi_i\frac{\sin [k(l_{i,i+1}-x)]}{\sin kl_{i,i+1}}+
\psi_{i+1}\frac{\sin kx}{\sin kl_{i,i+1}}\label{eq1}
\end{eqnarray}
\noindent
where, $\psi_i$ is the amplitude at the $i$th node, and $k$ is the absolute 
value of the wave vector in the loop. For an electromagnetic wave,
 $k=\frac{i\omega\sqrt{\epsilon}}{c_0}$ where $\omega$ and $c_0$ are the
 frequency and the speed (in vacuum) of the electromagnetic wave respectively.
 $\epsilon$ is the relative permittivity of the dielectric medium, which we may
 assume to be real. A complex dielectric constant can easily be dealt with. The
 continuity of the wave function at the nodes and the flux conservation 
criterion are used to map the problem of wave propagation in such loops into
 an equivalent tight binding problem of electron propagation on a one dimension
 lattice \cite{zqz94}.The resulting difference equation is,
\begin{equation}
(E-\epsilon_i)\psi_i=t_{i,i+1}\psi_{i+1}+t_{i-1,i}\psi_{i-1}\label{eq2}
\end{equation} 
\noindent
Here, we parametrize $E=2 \cos kl_{i,i+1}$ and, define an effective on-site
 potential 
\begin{eqnarray}
\epsilon_i & = & 2\cos kl_{i,i+1}+\sum_{m=1}^{N_{i,i-1}} \cot\theta_{i-1,i}^{(m)}+ \sum_{m=1}^{N_{i,i+1}} \cot\theta_{i,i+1}^{(m)}\label{eq3}
\end{eqnarray}
\noindent
and the effective nearest neighbour `hopping integral' is given by,
\begin{equation}
t_{i,i+1}=\sum_{m=1}^{N_{i,i+1}}\frac{1}{\sin\theta_{i,i+1}^{(m)}}\label{eq4}
\end{equation}
\noindent
where, $N_{i,i\pm1}$ is the number of segments between the nodes $i$ and
 $i\pm1$. $\theta_{i,i+1}^{(m)}=kl_{i,i+1}^{(m)}$ is the corresponding phase
 acquired in the $m$th segment. The problem of wave propagation now becomes
 mathematically equivalent to the problem of transmission of an electron with
 energy $E$ through the mapped lattice, where, $-2 \le E \le 2$.
\vskip .2in
It is now simple to eliminate the amplitudes $\psi_i$ corresponding to the
nodes denoted by the patterned circle in [Fig. \ref{fig1}(b)] in terms of the
 amplitudes at the remaining sites. The result is a Vicsek lattice of points
 [Fig. \ref{fig2}(a)]. In the equivalent language of the electronic problem,
we now have three distinct values for the on-site potentials, given by,
\begin{eqnarray}
\epsilon_A &=& 2 \cos \theta_3+ 2 \cot \theta_1+\cot \theta_2+\frac{4}{\sin^2\theta_1}
\left( \frac{1}{E- 2 \cos \theta_3- 2 \cot \theta_1-\cot \theta_2}\right) \nonumber \\
\epsilon_{B} &=& 2 \cos \theta_3+ \cot 2\theta_1+\cot \theta_2 \nonumber \\
\epsilon_C &=& 2 \cos \theta_3+ 4 \cot \theta_1+\frac{2}{\sin^2\theta_2}\left( \frac{1}
{E-2 \cos \theta_3-2 \cot \theta_1-\cot \theta_2}\right) \label{eq6}
\end{eqnarray}
and, the three nearest neighbour `hopping integrals' are given by:
\begin{eqnarray}
t_1 &=& \frac{2}{\sin\theta_1\sin\theta_2}\left( \frac{1}{E- 2\cos\theta_3-2\cot\theta_1
-\cot\theta_2}\right) \nonumber \\
t_2 &=& \frac{1}{\sin\theta_2} \nonumber \\
T & = & \frac{1}{\sin \theta_3}\label{eq7}
\end{eqnarray}
where, $T$ is the n.n. hopping integral between two neighbouring clusters. Here,
 $2a_1$, $a_2$, $a_3$ are the lengths of the loop, the wire and the connector
 connecting the two neighbouring clusters respectively. Therefore, 
$\theta_j=ka_j$ with $a_j=a_1,a_2,a_3$. This reconstruction facilates the use
 of RSRG to discern the extended and other modes of the lattice as well as to
compute the transmittivity. The results are now discussed.
\section{Unusual eigenmodes of a VWN}
\begin{figure}
\centering{\psfig{figure=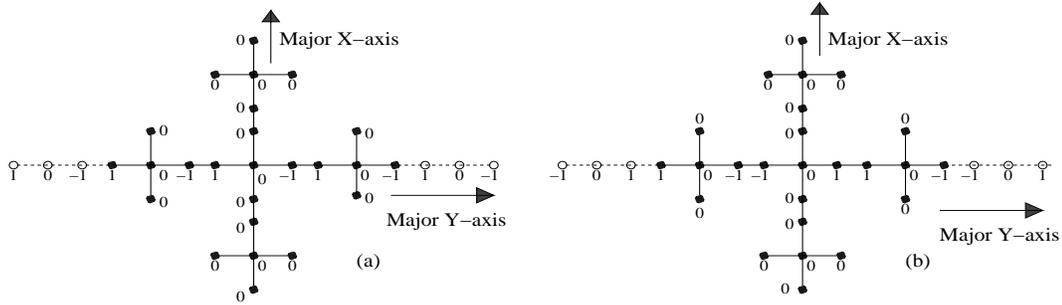,height=4cm,width=14cm}}
\caption{Distribution of amplitudes of the `extended' wavefunction for the lead
(dashed lines)-sample-lead systems corresponding to (a)$E=\epsilon_A-T$ and (b)
$E=\epsilon_A+T$. The transmission corresponding to these states across the
 sample decays with increasing system size as explained in the text.}
\label{fig3}
\end{figure}

Let us consider a VWN at any arbitrary generation (converted into a Vicsek
lattice, of points) clamped between two semi-infinite waveguide leads. Such a
construction in its second generation is shown in Fig \ref{fig3}. Now, a VWN 
at any generation $n+2$ can be renormalized $n$ times to obtain a second
 generation structure by decimating the appropriate vertices \cite{acb96}. The
 RSRG recursion relations for the on-site terms and the hopping integrals are 
given by,
\begin{eqnarray}
\epsilon_A(n+1) & = & \epsilon_A(n)+\frac{t_1^2(n)}{E-F_2(n)}
 \nonumber \\
\epsilon_B(n+1) & = & \epsilon_B(n)+\frac{t_2^2(n)}{E-F_2(n)}
 \nonumber \\
\epsilon_C(n+1) & = & \epsilon_C(n)+4 t_1^2(n)F_4(n) \nonumber \\
t_1(n+1) & = & \frac{t_1^3(n)T(n)F_1(n)}{[E-\epsilon_A(n)][E-F_2(n)]} 
\nonumber \\
t_2(n+1) & = & \frac{t_2(n)t_1^2(n)T(n)F_1(n)}{[E-\epsilon_A(n)][E-F_2(n)]}
 \nonumber \\
T(n+1) & = & T(n) \label{eq8}
\end{eqnarray}
\noindent
where, $n$ denotes the stage of renormalization and
\begin{eqnarray}
F_1(n) & = & \frac{E-\epsilon_A(n)}{[E-\epsilon_A(n)]^2-T^2(n)} \nonumber \\
F_2(n) & = & \epsilon_C(n)+\frac{2t_2^2(n)}{E-\epsilon_B(n)}+t_1^2(n)F_1(n)
 \nonumber \\
F_3(n) & = & \frac{E-\epsilon_B(n)}{[E-\epsilon_B(n)][E-\epsilon_C(n)]-2t_2^2(n)}
 \nonumber \\
F_4(n) & = & \frac{E-\epsilon_A(n)-t_1^2(n)F_3(n)}{[E-\epsilon_A(n)]
[E-\epsilon_A(n)-t_1^2(n)F_3(n)]-T^2(n)} \nonumber
\end{eqnarray}
\noindent
It is seen that $T$ remains fixed under RSRG iterations.
\vskip .2in
It is not difficult now to construct, by inspection, extended wave functions
 which have non trivial distribution of amplitudes along the X-axis. The wave
 vectors for these wavefunctions are obtained by setting
\begin{equation}
E=\epsilon_A(n)\pm T \label{eq9}
\end{equation}
\noindent
As the RSRG process can proceed, in principle indefinitely, one can extract an
infinity of different $k$-values by solving the equation (\ref{eq9}). A couple
 of such states are illustrated in Fig. \ref{fig3} on a second stage fractal
 for $E=\epsilon(n)\pm T$. It should be noted that $\epsilon_C$ in general
 flows to infinity with the progress of re-normalization. Therefore, it is
 necessary to choose the amplitudes at the central site of all the basic
 five-site cells to be zero in order to avoid the divergence of amplitudes at
 any arbitrary site. The `extended' character of these eigenmodes are usually
 tested from a study of the flow of the `hopping integrals' under successive
 RSRG steps.
\vskip .2in
The recursion relations (\ref{eq8}) reveal that if, $E=\epsilon_A(n)-T$ at any
 $n$th stage of re-normalization, then $\epsilon_A$, $\epsilon_B$, $t_1$ and
 $t_2$ immediately reach their fixed points i.e.
 $\epsilon_A(n+1)=\epsilon_A(n)$, $\epsilon_B(n+1)=\epsilon_B(n)$,
 $t_1(n+1)=t_1(n)$ and $t_2(n+1)=t_2(n)$ for all subsequent stages of
 renormalization. This implies that at any scale of length there is a nonzero
 overlap between the wave-functions at the nearest neighbour sites (at that
 scale), and that the amplitudes remain finite everywhere in even an infinite
 lattice. For example, solving $E=\epsilon_A-T$ we get $ka=0.547457038567240$
 and $2.118253365362137$ with $a_1=a_2=a_3=a=1$. The hopping integrals $t_1$
 and $t_2$ remain fixed at nonzero values (corresponding to their fixed points)
 in these two cases. Thus, from an RSRG point of view we expect an extended
 character of such states. To see whether such states are `transparent' in
regard of the incoming wave, we have to compute the transmission of an
electromagnetic wave incident on a finite VFN through leads connected to the
 two-ends of the fractal lattice.
\section{Transmission coefficient}
\subsection{Formulation}
To calculate the transmission coefficient we first renormalize the Vicsek
lattice $n$-times using the RSRG decimation method. The result is a Vicsek
lattice consisting of five vertices and characterized by the $n$-times
renormalized values of the `on-site' potentials and the `hopping integrals'.
The recursion relations are given by Eq. (\ref{eq8}). This renormalized version
 [Fig. \ref{fig2}(b)] is now reduced to a 3-site cluster [Fig. \ref{fig2}(c)]
 and the effective on-site potential of the central site becomes
\begin{eqnarray}
\tilde\epsilon_C(n) & = & \epsilon_C(n)+\frac{2t_2^2(n)}{E-\epsilon_B(n)}
\label{eq10}
\end{eqnarray}
\noindent
\begin{figure}
\centering{\psfig{figure=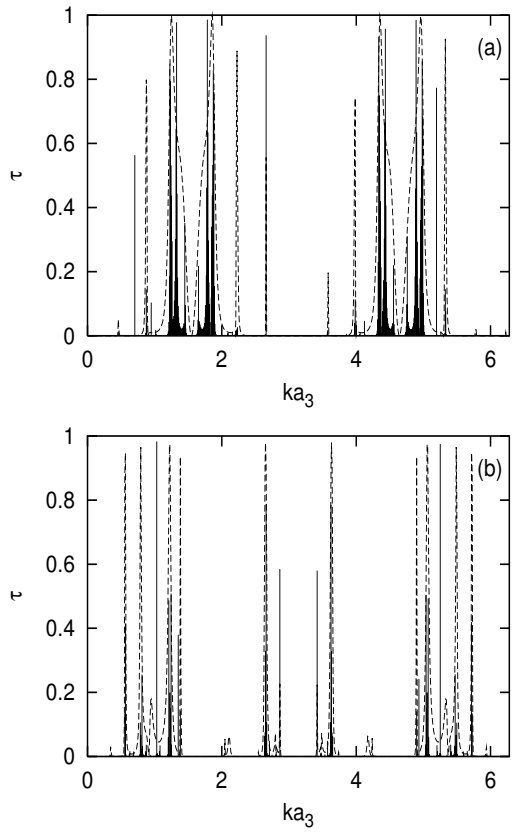,height=16cm,width=10cm}}
\caption{(a) $\tau$ vs $ka_3$ with $a_1=0.01$, $a_2=a_3=1$. (b) $a_1=0.05$,
$a_2=a_3=1$. The solid and the dashed lines in each figure indicate results
for the second and the first generation VWN respectively.}\label{fig4}
\end{figure}
This triatomic `molecule' is now clamped between two semi-infinite perfect
waveguides, which are represented, following the discretization process 
outlined in Ref \cite{zqz94}, by a sequence of identical site potentials
\begin{equation}
\epsilon_0 = 2\cos \theta_3+2\cot \theta_3
\label{eq11}
\end{equation}
\noindent
and nearest neighbour hopping integral
\begin{equation}
t_0=\frac{1}{\sin \theta_3}
\label{eq12}
\end{equation}
The decimation procedure is illustrated in Fig. \ref{fig2} for a second
generation fractal. The equations are easily obtained by artificially
 considering the lead to be consisting of identical materials of length $a_3$
 joined end-to-end. The transmission coefficient $\tau$ is now given by
 \cite{ads81},
\begin{equation}
\tau=\frac{4\sin ^2 \theta_3}{[P_{12}-P_{21}+(P_{11}-P_{22})\cos \theta_3]^2+
(P_{11}+P_{22})^2 \sin ^2\theta_3}\label{eq13}
\end{equation}
\noindent
where,
\begin{eqnarray}
P_{11} & = & \frac{[E-\epsilon_A(n)]^2[E-\tilde\epsilon_C(n)]}{t_1^2(n)t_0}-
\frac{2[E-\epsilon_A(n)]}{t_0}\nonumber \\
P_{12} & = & 1-\frac{[E-\epsilon_A(n)][E-\tilde\epsilon_C(n)]}{t_1^2(n)} 
\nonumber \\
P_{21} & = & - P_{12} \nonumber \\
P_{22} & = & -\frac{t_0[E-\tilde\epsilon_C(n)]}{t_1^2(n)} \nonumber
\end{eqnarray}
\subsection{Results}
{\bf (a)} Let us analyze the wavefunction which corresponds to an extendedd
 state of the lead-VWN-lead system. For the values of wave vector correspond
 to the energy $E=\epsilon_A-T$, for example, we can use the recursion
 relations (\ref{eq8}) to write explicitly the renormalized values of 
$\epsilon_C$ in the successive generations as,
\begin{displaymath}
\epsilon_c(1)=5\epsilon_c-\sigma
\end{displaymath}
\begin{displaymath}
\epsilon_c(2)=5^2\epsilon_c-5\sigma-\sigma
\end{displaymath}
\noindent
and so on. Here
\begin{displaymath}
\sigma=4(\epsilon_A-T)-\frac{8t_2^2}{\epsilon_A-\epsilon_B-T}+\frac{4t_1^2}{T}
\end{displaymath}
\noindent
is constant because $\epsilon_A$, $\epsilon_B$, $t_1$, $t_2$ and $T$ attain 
 their fixed point values for $n\ge 1$. Continuing in this way we get, after
 $n$ steps of renormalization,
\begin{displaymath}
\epsilon_c(n)=5^n\epsilon_c-P_{n-1}(\sigma)
\end{displaymath}
\noindent
where,
\begin{displaymath}
P_{n-1}(\sigma)=5^{n-1}\sigma+5^{n-2}\sigma+...........+\sigma
\end{displaymath}
\noindent
and $5^n=N$, the system size of the lattice at its $n$th generation.
 Therefore, we see that for large values of $n$, i.e., for large size of the
 original lattice $\epsilon_C(n)\sim N$ as $\epsilon_A$, $\epsilon_B$, $t_1$
 and $t_2$ are invariant at each step of renormalization. So for large $n$,
 its simple to work out from Eq. (\ref{eq9}) that $\tau\sim \frac{1}{N^2}$ 
indicating power law decay as a function of the total number of matrices. This
 is incompatible with the invariance of the hopping integrals under RSRG. The
 growth of $\epsilon_C(n)$ under RSRG implies that an electromagnetic wave
 travelling through the perfect waveguide leads and entering the fractal
 structure will experience larger and larger effective `potentials' at the
 centres of the five site clusters. For large enough VWN systems we thus have
 the possibility of a power law localization of light.
\vskip .2in
The existence of the dangling side units consisting of a wire and a loop which
 themselves form a fractal geometry, causes destructive interference between
the waves propagating and reflecting back and forth along these side units. As
 for all these wave vectors the amplitude of the wave function is pinned at the
 value zero at the centerers of five site clusters at different length scale,
 standing waves are formed along the branches with nodes at the centers of one 
or the other five site cluster, which should localize the light wave and reduce
 the transmission. We call such modes `unusual', and emphasize that these are
eigenmodes not for a finite cluster, but for the {\it lead-VWN-lead system},
 irrespective of its size.
\begin{figure}
\centering{\psfig{figure=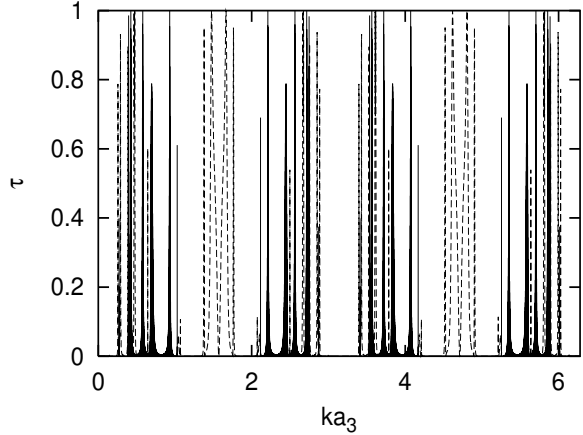,height=6cm,width=8cm}}
\caption{$\tau$ vs $ka_3$ with $a_1=2$, $a_2=a_3=1$. The solid and the dashed 
lines now represent the results for the first and the second generation VWN
 respectively. The appearance of new transmission zones is clearly seen between
 $ka_3=0$ and $2$, $4$ and $6$.}\label{fig5}
\end{figure}
\vskip .2in
{\bf (b)} We have also examined the features of the transmission spectrum of
 finite VWN's of arbitrary size. The appearance of transmission maxima
 separated by photonic gaps is a general phenomenon. The fragmentation in the
 spectrum increasing with increasing size of the system. Figures (\ref{fig4})
 and (\ref{fig5}) exhibit the main characteristics where, in addition to the
 variation of $\tau$ against the wave vector, we study the effect of the change
of the size of the loops ($a_1$). Three cases have been explicitly shown with
 the size of the loop ($a_1$) increasing from a very small value to a value
 comparable to the lengths of the other waveguide segments ($a_2$ and $a_3$).
\vskip .2in
Though in every case, the exact distribution of values of the $tau$ is 
sensitive to the individual values of the parameters, still some interesting
 behaviour is observed in certain cases. For example, setting $a_2=a_3=1$
 and $a_1=0.01$ in arbitrary units, the transmission spectrum shows the
 presence of clusters of high transmittivity flanked by sharp delta-like peaks,
 even for a first generation VWN [dashed line in Fig. \ref{fig4}(a)]. The same
 set of parameters leads to a fragmented band structure as we move over to the
 second generation [vertical  line in Fig. \ref{fig4}(b)]. The spectrum reveals
 almost similar features as $a_1$ is gradually increased from $0.01$ to higher
 values. For $a_1=0.5$, we get a clear signature of wider photonic gaps
 separating four clusters of high transmittivity. This is seen in Fig. 
\ref{fig4}(b), where the dashed and the solid lines again correspond to the
 first and second stage VWN.
\begin{figure}
\centering{\psfig{figure=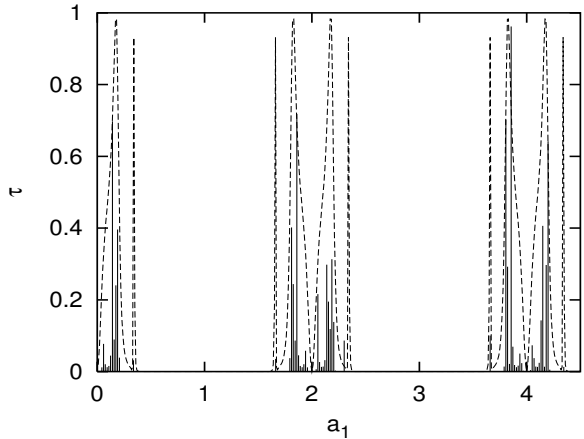,height=6cm,width=8cm}}
\caption{$\tau$ vs $a_1$ with $a_2=a_3=1$ and $k=\pi/2$.}\label{fig6}
\end{figure}
\vskip .2in
When the dimension of the loop starts getting larger than the connecting
 waveguide segments, we find that new clusters of transmission maxima start
 appearing in the major photonic gaps of the earlier generation 
[Fig. \ref{fig5}]. The three major photonic gaps corresponding to the first
 generation fractal get populated by two major photonic bands (dashed line in
Fig. \ref{fig5}) when the length of the loop exceeds the other two segments 
$a_2$ and $a_3$. 
\vskip .2in
To make the appearance of the photonic gaps more pronounced, in Fig. \ref{fig6}
we show the variation of transmission coefficient $\tau$ of a VWN in the first
 (dashed curve) and the second (solid vertical lines) respectively. The more or
 less fixed zones of zero transmission around $a_1=1$ and $a_1=3$ mark the
 broad photonic band gaps in a VWN.
\section{Conclusion}
We have shown that light transmission in a Vicsek fractal geometry is marked
by the appearance of unusual extended-like eigenmodes for which a non-trivial 
distribution of finite amplitudes can be obtained in a Vicsek waveguide network
 clamped between semi-infinite leads. The transmittivity of such a lead-network
-lead system shows a power law decay as the network grows bigger. Broad
photonic gaps are obtained by appropriately choosing the network parameters.
\vskip .5in
\begin{center}
{\bf Acknowledgment}
\end{center}
\vskip .2in
AC acknowledge CSIR, India for financial help (grant no. 03(0944)/02/EMR-II)
and SS thanks UGC India for financial help through a research fellowship.

\end{document}